\documentclass[twocolumn,showpacs,showkeys]
{revtex4}

\usepackage{latexsym}
\usepackage{bm}

\def\cS{{\cal S}}
\def\cR{{\cal R}}
\def\Tr{{\rm Tr}}
\def\+-{\buildrel + \over -}
\def\ket#1{\mid~\!\!\!{#1}~\!\!\rangle}
\def\bra#1{\langle~\!\!{#1}~\!\!\!\mid}

\begin{document}

\title{Chains of Quasi-Classical
Informations\\ for Bipartite
Correlations and the Role of Twin
Observables}

\author{Fedor Herbut}
\affiliation {Faculty of Physics,
University of Belgrade, POB 368,
Belgrade 11001, Serbia; also Serbian
Academy of Sciences and Arts, Knez
Mihajlova 35, 11000 Belgrade}

\email{fedorh@infosky.net}

\date{\today}

\begin{abstract}
Having the quantum correlations in a
general mixed or pure bipartite state
in mind, the part of information
accessible by simultaneous measurement
on both subsystems is shown {\it never
to exceed} the part accessible by
measurement on one subsystem, which, in
turn is proved {\it not to exceed} the
von Neumann mutual information. A
particular pair of (opposite-subsystem)
observables are shown to be responsible
both for the amount of quasi-classical
correlations and for that of the purely
quantum entanglement in the {\it
pure-state} case: the former via
simultaneous subsystem measurements,
and the latter through {\it the entropy
of coherence or of incompatibility},
which is defined for the general case.
The observables at issue are so-called
{\it twin observables}. A general
definition of the latter is given in
terms of their detailed properties.
\end{abstract}



\maketitle

\rm As it is well known, quantum
information theory is closely connected
with the correlations inherent in an
{\it arbitrary} bipartite state
(mathematically: statistical operator)
$\rho_{12}$ of a composite system
$1+2$. The correlations have
surprisingly many facets, and the
relations among them are the subject of
intense current investigation. This
article is intended to make a
contribution to the issue.

Let us define some quantitative
elements of correlations. The subsystem
states (reduced statistical operators)
are $\rho_s\equiv
\Tr_{s'}\rho_{12},\quad
s,s'=1,2;\enskip s\not= s'$
("$\Tr_{s'}$" is a partial trace), and
we have the three {\it von Neumann
entropies}: $S(n)\equiv S(\rho_n)\equiv
-\Tr_n(\rho_n log\rho_n),\quad
n=1,2,12$. One of the basic correlation
or entanglement entities is the {\it
von Neumann mutual information}:
$$I(1:2) \equiv
S(1)+S(2)-S(12).\eqno{(1)}$$ It is
conjectured that it is the amount of
{\it total} correlations \cite{Ved}.

For the purpose of notation, let us
write down an arbitrary first-subsystem
and an arbitrary second subsystem
complete observable (Hermitian
operator) with purely discrete spectra:
$ A_1=\sum_ia_i\ket{i}_1\bra{i}_1,
\quad B_2=\sum_jb_j\ket{j}_2\bra{j}_2.$
The {\it measurement} of $A_1\otimes 1$
gives rise to the distant (as opposed
to "direct") state decomposition $
\rho_2=\sum_ip_i\rho^i_2,$ where $
p_i\equiv \Tr [\rho_{12}
(\ket{i}_1\bra{i}_1\otimes 1)] $ is the
probability of the result $a_i$, and $
\rho^i_2\equiv p_i^{-1}
\Tr_1[\rho_{12}(\ket{i}_1\bra{i}_1\otimes
1)] $ is the opposite-subsystem state
corresponding to this result if
$p_i>0$.

Entropy is concave \cite{Wehrl}
(section II.B there), i. e.,
$\sum_ip_iS(\rho^i_2)\leq S(2)$, and $$
I(m1\rightarrow 2)_A\equiv
S(2)-\sum_ip_iS(\rho^i_2)\eqno{(2a)}$$
is the {\it information gain} about
subsystem $2$ on account of the direct
{\it measurement} of the observable
$A_1$ on subsystem $1$. Symmetrically,
one defines the symmetric quantity
$I(1\leftarrow m2)_B$.

One, further, defines \cite{Ved},
\cite{Zur} $$ I(m1\rightarrow 2)\equiv
sup\{I(m1\rightarrow 2)_A\},
\eqno{(2b)}$$ the {\it supremum} taken
over all complete $A_1$, as the largest
amount of information (contained in the
correlations) accessible by measurement
of an observable on the first
subsystem. Symmetrically, one defines
the symmetric quantity $I(1\leftarrow
m2)\equiv sup\{I(1\leftarrow m2)_B\}$
over all second-subsystem complete
measurements.

If one performs simultaneous
measurement of $(A_1\otimes 1)$ and of
$(1\otimes B_2)$ on $\rho_{12}$
(denoted by $(A_1\wedge B_2)$), then
one deals with a classical discrete
joint probability distribution
$p_{ij}\equiv \Tr [\rho_{12}
(\ket{i}_1\bra{i}_1\otimes
\ket{j}_2\bra{j}_2)]$. It implies, in
its turn, the {\it mutual information}
$I(m1:m2)_{A\wedge B}$ via the Gibbs-
Boltzmann-Shannon entropies $H(A,B)
\equiv -\sum_{ij}p_{ij}logp_{ij}$,
$H(A)\equiv -\sum_ip_ilogp_i$, $H(B)
\equiv -\sum_jp_jlogp_j$, where $p_i
\equiv \sum_jp_{ij}$ and $p_j\equiv
\sum_ip_{ij}$ are the marginal
probability distributions. Then $$
I(m1:m2)_{A\wedge B}\equiv H(A)+H(B)
-H(A,B).\eqno{(3a)}$$

Finally, $$ I(m1:m2)\equiv
sup\{I(m1:m2)_{A\wedge B}\}
\eqno{(3b)}$$ over all choices of
complete observables $A_1$ and $B_2$.
This is the largest amount of
information  on a subsystem observable
(contained in the quantum correlations)
accessible by measurement of an
observable on the opposite subsystem.

{\it The claimed chains of information
inequalities, valid for every bipartite
state} $\rho_{12}$, go as follows: $$
0\leq I(m1:m2)\leq I(m1\rightarrow
2)\leq min\{I(1:2),S(2)\},\eqno{(4a)}$$
$$0\leq I(m1:m2)\leq I(1\leftarrow
m2)\leq
min\{I(1:2),S(1)\}.\eqno{(4b)}$$
 Both in (4a) and in (4b) one has
equality in the first inequality {\it
if and only if} the state $\rho_{12}$
is {\it uncorrelated}, i. e.,
$\rho_{12}= \rho_1\otimes \rho_2$.

The role of $S(2)$ in the last
inequality in (4a) is obvious from
(2a), and symmetrically for (4b).

In the classical discrete case both
chains (4a) and (4b) contain only
equalities, and one has $I(1:2)\leq
S(1),S(2)$. As to the corresponding
inequality in the quantum case, one
cannot do better than $I(1:2)\leq
2S(1),2S(2)$ \cite{Lieb75}.

The inequality $I(m1:m2)\leq I(1:2)$
implied by (4a), together with the
stated necessary and sufficient
condition for equality in the first
inequality in (4a), was proved in 1973
by Lindblad \cite{Lin73} (Theorem 2.
there). The third inequality in (4b)
with $I(1:2)$ was claimed and a proof
was presented in \cite{Zur}. (It is
perhaps useful to have an independent
verification like the one in this
article.)

The second inequality in (4a) is being
proved for the first time in this
article I believe. For the sake of
completeness, let me prove the entire
chain.

The inequalities in (4a) are,
essentially, {\it a consequence} of a
{\it result of Lindblad} of classical
value \cite{Lin4} (see Corollary
there), and (4b) follows symmetrically.
To explain this claim, let me introduce
the so-called {\it relative entropy} of
a quantum state (statistical operator)
$\sigma$ with relation to another state
(statistical operator) $\rho$: $$
S(\sigma |\rho)\equiv \Tr \sigma log
\sigma -\Tr \sigma log\rho. $$
 One has $0\leq
S(\sigma |\rho )$ with equality if and
only if $\sigma =\rho$.

Lindblad's result involves the {\it
ideal measurement} of an arbitrary
complete or incomplete observable
(Hermitian operator) $A$ with a purely
discrete spectrum. Let its unique
spectral form, i. e., the one without
repetition in the characteristic
values, be $A=\sum_ia_iP^i$. Denoting
by $T_A\sigma $ the state into which
$\sigma$ changes due to the
nonselective ideal measurement of $A$
in it, one has $$ T_A
\sigma=\sum_iP^i\sigma P^i\eqno{(5)} $$
\cite{Lud}, and {\it Lindblad's result
states} that $$ S(T_A\sigma |T_A\rho
)\leq S(\sigma |\rho ). \eqno{(6)}$$

One should note that also the RHS of
(5) is a statistical operator. Hence,
for any other observable $B=\sum_jb_j
Q^j$, (6) implies, what may be called,
{\it the Lindblad chain} $S(T_BT_A
\sigma |T_BT_A\rho )\leq S(T_A\sigma
|T_A\rho )\leq S(\sigma |\rho )$. One
may even extend the measurements to
operations \cite{Lin75}.

{\it The von Neumann mutual
information} in any bipartite state
$\rho_{12}$ can be expressed {\it in
terms of relative entropy}: $$ I(1:2)=
S(\rho_{12}|\rho_1\otimes \rho_2).
\eqno{(7)}$$ (This known claim is
easily checked utilizing $log(\rho_1
\otimes \rho_2)=(log\rho_1)\otimes 1+
1\otimes (log\rho_2)$, which, in turn,
is easily seen in spectral forms.)

I am going to demonstrate that the
claimed chain of inequalities (4a) is a
consequence of the Lindblad chain: $$
0\leq
S\Big(T_AT_B\rho_{12}|T_AT_B(\rho_1\otimes
\rho_2)\Big)\leq$$ $$\leq
S\Big(T_A\rho_{12}|T_A (\rho_1\otimes
\rho_2)\Big)\leq S(
\rho_{12}|\rho_1\otimes \rho_2)
\eqno{(8)}$$ with subsystem observables
$A_1$ and $B_2$ that are {\it complete}
in some subspaces $\cS_1$ and $\cS_2$
containing the ranges of the operators
$\rho_1$ and $\rho_2$ respectively.

In order to recognize the meaning of
{\it the first relative entropy} in
(8), we make use of the auxiliary claim
that for complete or incomplete
subsystem observables
$A_1=\sum_ia_iP_1^i$ and
$B_2=\sum_jb_jQ_2^j$ (unique spectral
forms), and for any bipartite state
$\rho_{12}$, one has $$
T_A\rho_1=\Tr_2(T_AT_B\rho_{12}), \quad
T_B\rho_2=\Tr_1(T_AT_B\rho_{12}),
\eqno{(9)}$$ where $\rho_s$, $s=1,2$,
are the subsystem states of
$\rho_{12}$. Relations (9) are proved
in Appendix 1.

Further, one can argue with Lindblad
\cite{Lin73} (Theorem 2. there) as
follows. Making use of (1), (7) and
(9), one obtains
$S\Big(T_AT_B\rho_{12}|T_AT_B(\rho_1
\otimes \rho_2)\Big)=S(T_A\rho_1)
+S(T_B\rho_2)-S(T_AT_B\rho_{12}).$
Taking $T_A$ and $T_B$ in explicit form
(cf (5) {\it mutatis mutandis}), we see
that we have a mixture of {\it
orthogonal} pure states. The so-called
{\it mixing property} of entropy allows
us to write it as the sum of the
so-called {\it mixing entropy} (that of
the statistical weights) and the
average entropy \cite{Wehrl} (see
section II.F. and II.B. there). Since
pure states have zero entropy, one
obtains:
$$LHS=H(A)+H(B)-H(A,B)=I(m1:m2)_{A\wedge
B}.$$

Next, we turn to {\it the second
relative entropy} in (8). Utilizing
again relations (9) (this time with
$B\equiv 1$), (7) and (1), one obtains
$$S\Big(T_A\rho_{12}|T_A (\rho_1\otimes
\rho_2)\Big)=S(T_A\rho_1)+S(\rho_2)
-S(T_A\rho_{12}).\eqno{(10)}$$ Since
$A_1=\sum_i a_i\ket{i}_1\bra{i}_1$, for
$p_i\equiv \Tr
(\ket{i}_1\bra{i}_1\rho_{12})>0$, one
has
$$\ket{i}_1\bra{i}_1\rho_{12}\ket{i}_1
\bra{i}_1=p_i\ket{i}_1\bra{i}_1\otimes
\rho_2^i,\eqno{(11a)}$$  $$
\rho_2^i\equiv p_i^{-1}
\Tr_1\Big(\ket{i}_1\bra{i}_1\rho_{12}
\ket{i}_1 \bra{i}_1\Big).\eqno{(11b)}$$
(The tensor factor "$\otimes 1$" is
repeatedly omitted because no confusion
can arise.) The validity of (11a) is
straightforward to check in any pair of
orthonormal and complete subsystem
bases.

On account of (11a) and the fact that
both $T_A\rho_1$ and $T_A\rho_{12}$ are
orthogonal mixtures of states (cf (5))
with the same statistical weights, we
can apply the mixing property of
entropy both to $S(T_A\rho_1)$ and to
$S(T_A\rho_{12})$. Then, the LHS of
(10) becomes equal to
$$H(A)+S(\rho_2)-\Big(H(A)
+\sum_ip_iS(\rho_2^i)\Big)=I(m1\rightarrow
2)_A$$ (cf (2a)).

The chain (8) can now be rewritten as
$$0\leq I(m1:m2)_{A\wedge B}\leq$$
$$\leq I(m1\rightarrow 2)_A\leq
I(1:2).\eqno{(12)}$$ (The symmetric
chain is derived symmetrically.)

The inequality $$I(m1:m2)_{A\wedge
B}\leq I(m1\rightarrow 2)_A$$ has the
obvious {\it physical interpretation}
that, in general, only part of the
quantum information gain about
subsystem $2$ due to the measurement of
$A_1$ can be realized as information
about a concrete complete observable
$B_2$.

The same inequality implies that the
{\it quantum information gain}
$I(m1\rightarrow 2)_A$ is an upper
bound to any concrete information
$I(m1:m2)_{A\wedge B}$ about some
$B_2$.

Taking the {\it suprema} in (12), and
having (2a) in mind, one obtains (4a).

In \cite{Ved} $I(m1\rightarrow 2)$ is
interpreted as the quasi-classical part
of the amount of quantum correlations
in any bipartite state $\rho_{12}$. The
authors define the so-called relative
entropy of entanglement
$E_{RE}(\rho_{12})\equiv
inf\{S(\rho_{12}|\sigma_{12})\}$, where
the infimum is taken over all separable
states $\sigma_{12}$, as a measure of
(purely quantum) entanglement (cf also
\cite{Ved'}). Since $I(1:2)=
S(\rho_{12}|\rho_1\otimes \rho_2)$,
obviously, $E_{RE}(\rho_{12})\leq
I(1:2)$.

Essentially the same view of
$I(m2\rightarrow 1)$ as in \cite{Ved}
is, independently, taken in \cite{Zur}.
The latter authors call the difference
$$\delta(m2\rightarrow 1)\equiv
I(1:2)-I(m2\rightarrow 1)\eqno{(13)}$$
{\it "quantum discord"}, and they
interpret it as the truly quantum part
of the total amount of correlations
$I(1:2)$. (It is inaccessible to
subsystem measurement.)

Next we apply the derived chain of
quasi-classical informations to pure
states. They represent a simple enough
case to gain detailed insight.\\

{\it Quasi-classical informations in
bipartite pure states.-} We turn now to
a general {\it pure state}
$\rho_{12}\equiv \ket{\Phi}_{12}
\bra{\Phi}_{12}$. Let us write
$\ket{\Phi}_{12}$ as a Schmidt
decomposition \cite{Per}, \cite{FM76}
into biorthogonal state vectors:
$$\ket{\Phi}_{12}=\sum_ir_i^{1/2}
\ket{i}_1\ket{i}_2.\eqno{(14)}$$ Taking
$$A_1\equiv \sum_ia_i\ket{i}_1
\bra{i}_1,\quad 0\not= a_i\not= a_{i'}
\not= 0\enskip \mbox{for}\enskip i\not=
i',\eqno{(15a)}$$ $$B_2\equiv
\sum_ib_i\ket{i}_2\bra{i}_2,\quad
0\not= b_i\not= b_{i'} \not= 0\enskip
\mbox{for}\enskip i\not=
i',\eqno{(15b)}$$ one obtains for the
induced classical discrete probability
distribution (cf (3a)): $p_{ij}=
\delta_{ij}r_i$. Then
$$I(m1:m2)_{A\wedge
B}=H(A)=H(B)=H(A,B)=S(1)$$
$$=S(2)=I(m1\rightarrow 2)=
I(1\leftarrow m2)\eqno{(16)}$$ (cf (4a)
and (4b) without $I(1:2)$). It is seen
from (3b) that $I(m1:m2)_{A\wedge B}$
is a lower bound to all quantities in
the chains (4a) and (4b), and it
reaches its highest possible value
$S(1)=S(2)$ in $\ket{\Phi}_{12}$ (cf
(16)). Hence, it equals not only
$I(m1:m2)$, but also $I(m1\rightarrow
2)$ and $I(1\leftarrow m2)$.

Besides, also $$\delta(m1\rightarrow
2)= \delta(1\leftarrow m2)
=S(1)=S(2)\eqno{(17)}$$ (because
$I(1:2)=2S(1)=2S(2)$). The same
quantity, called entropy of
entanglement and denoted by
$E(\ket{\Phi}_{12})$ was obtained in
\cite{PopR}.

Returning to the above quasi-classical
informations in $\ket{\Phi}_{12}$, one
can say that the pair $(A_1,B_2)$ of
opposite subsystem observables (15a)
and (15b) actually {\it realize, in
simultaneous measurement, the entire
part of the total correlations that is
available for subsystem measurement}.
This pair of observables has noteworthy
properties. Next, we resort to a
sketchy presentation of them in the
general case.\\

{\it Twin observables with respect to a
general bipartite state}.

Let us now turn to a concise but
sufficiently detailed definition of
{\it twin observables}, which is wider
than the one given in previous work
\cite{FM76}, \cite{twins}. All
necessary proofs are provided in
Appendix 2.

Let $\rho_{12}$ be an arbitrary given
bipartite state, and let $A_1$ and
$B_2$ be opposite-subsystem observables
(Hermitian operators) having the
following {\it three properties} with
respect to $\rho_{12}$:

(i) The operators {\it commute} with
the corresponding reduced statistical
operators: $[A_1,\rho_1]=0,\quad
[B_2,\rho_2]=0$.

On account of the commutations, the
(topological closures $\bar
\cR(\rho_i)$ of the) ranges
$\cR(\rho_i)$, $i=1,2$, are {\it
invariant} subspaces for $A_1$ and
$B_2$ respectively, and the operators
have {\it purely discrete spectra} in
them. These are precisely the {\it
detectable} parts of the respective
spectra of $A_1$ and $B_2$, i. e., they
consist of those characteristic values
that have positive probability in
$\rho_{12}$.

(ii) The detectable parts of the
spectra of $A_1$ and $B_2$ consist of
{\it an equal number} of characteristic
values, i. e., they are of the same
power.

(iii) One can establish a one-to-one
map between the two detectable parts of
the spectra {\it such that} the
corresponding characteristic values,
denoted by the same index $i$, {\it
satisfy for all value of} $i$ one of
the following four conditions:

 (a) {\it The information-theoretic condition}:
$$p_{ii'}\equiv \Tr
\rho_{12}P_1^iP_2^{i'}=\delta_{i,i'}
p_i,$$ where $P_1^i$ is the
characteristic projector of $A_1$
corresponding to the detectable
characteristic value $a_i$ and
symmetrically for $P_2^{i'}$ and
$b_{i'}$ of $B_2$; and $p_i\equiv \Tr
\rho_1P_1^i$ is the probability of
$P_1^i$ in $\rho_{12}$.

(b) {\it The measurement-theoretic
condition}: $$P_1^i\rho_{12}P_1^i=
P_2^i\rho_{12}P_2^i.$$

(c) {\it The condition in terms of
quantum logic}: $$\Tr
[\rho_2(P_1^i)P_2^i]=1,$$ where
$\rho_2(P_1^i)\equiv p_i^{-1}
\Tr_1\rho_{12}P_1^i$ is {\it the
conditional state} of subsystem $2$
when the event $P_1^i$ occurs.

(d) {\it The algebraic condition}:
$$P_1^i\rho_{12}=P_2^i\rho_{12}.$$

The four conditions in property (iii)
are {\it equivalent}.

If $A_1$ and $B_2$ do have the
mentioned three properties, then we
call them {\it twin observables for}
$\rho_{12}$. If all characteristic
values of $A_1$ and $B_2$ in $\bar
\cR(\rho_1)$ and $\bar \cR(\rho_2)$
respectively are {\it nondegenerate},
i. e., if $\forall i:\enskip \Tr P_s^i
Q_s=1$, where $Q_s$ is the range
projector of $\rho_s$, $s=1,2$, we say
that $A_1$ and $B_2$ are {\it complete}
twin observables with respect to
$\rho_{12}$.

{\it Comments} on the four conditions
in property (iii).

(a) The probability distribution
$p_{ii'}=\delta_{i,i'}p_i$ is the best
possible classical information channel:
a so-called lossless and noiseless one.
It is obvious that the correspondence
between the detectable parts of the
spectra is {\it unique}.

(b) The detectable characteristic
values $a_i$ of $A_1$ and $b_i$ of
$B_2$ are {\it equally probable} in
$\rho_{12}$. Besides, the ideal
measurement of $A_1$ and that of $B_2$
(actually of ($A_1\otimes 1_2$) and of
($1_1\otimes B_2$)) convert $\rho_{12}$
into the same state (cf the general
formula of L\"{u}ders for ideal
measurement \cite{Lud}). This makes
possible so-called {\it distant
measurement} \cite{FM76}: One can
measure $B_2$ in $\rho_{12}$ without
any dynamical influence on the second
subsystem by just measuring $A_1$ on
the first subsystem (or vice versa) in
the state $\rho_{12}$ of the bipartite
system.

(c) For an arbitrary event (projector)
$E_2$ for subsystem $2$ one can write
$$\Tr [\rho_{12}P_1^iE_2]=p_i
\Tr[\rho_2(P_1^i)E_2],$$ i. e., one can
factorize coincidence probability into
probability of the condition $P_1^i$
and conditional probability of the
event $E_2$ (in analogy with classical
physics). The conditional state
$\rho_2(P_1^i)$, when giving
probability one, {\it extends} the
absolute implication in quantum logic
(which is $E\leq F\enskip
\Leftrightarrow \enskip EF=E$, $E$ and
$F$ projectors) by {\it state-dependent
implication} \cite{impl}. This makes
$P_1^i$ and $P_2^i$ to imply each other
$\rho_{12}$-dependently.

(d) Since the detectable characteristic
values of twin observables $A_1$ and
$B_2$ are arbitrary, one can choose
them equal: $\forall i:\enskip
a_i=b_i$. Then the algebraic condition
{\it strengthens} into $$A_1\rho_{12}=
B_2\rho_{12}.$$

This case was studied in detail in
previous work \cite{FM76},
\cite{twins}. It was shown that the
stronger algebraic condition implies
all three above properties, i. e., that
it by itself makes $A_1$ and $B_2$ twin
observables (as defined in this
article) with the additional property
(iv): $\forall i:\enskip a_i=b_i$. It
was also shown that in the pure state
case the multiplicities of $a_i$ and
$b_i$ necessarily coincide, but they
need not be equal in the mixed-state
case.

 Without property (iv) twin
observables have a wider scope of
potential application.

Let us return to the above discussion
of quasi-classical informations
inherent in a given pure state vector
$\ket{\Phi}_{12}$. In view of the
information-theoretic condition in
property (iii) of twin observables, it
clearly follows from the above
discussion of (15a) and (15b) that one
is dealing with twin observables.

One can say that it is the pair
$(A_1,B_2)$ of twin observables given
by (15a) and (15b) that realizes, in
simultaneous measurement, the entire
quasi-classical information.

The ideal nonselective measurements of
$A_1$, that of $B_2$, and that of
$A_1\wedge B_2$ each convert
$\ket{\Phi}_{12}$ into one and the same
mixed state $$\rho_{12}'\equiv
\sum_ir_i\ket{i}_1\bra{i}_1\otimes
\ket{i}_2\bra{i}_2\eqno{(18)}$$ (cf
(14)).

As it is easily seen, the same pair of
observables (15a) and (15b) are {\it
complete twin observables} not only
with respect to $\ket{\Phi}_{12}$, but
also regarding $\rho'_{12}$. Also (16)
holds true for the latter. Again, the
same pair of twin observables "carry"
the entire
subsystem-measurement-accessible part
of information. But instead of (17), we
have zero quantum discord. There is no
subsystem-measurement-inaccessible part
of information. (No wonder, we are
dealing with a biorthogonal separable
mixed state in (18).)

In view of the fact that twin
observables have a variety of
particular properties, one may wonder
if the pair given by (15a) and (15b)
is, perhaps, of some relevance also for
the quantum discord in
$\ket{\Phi}_{12}$ (cf (14)). To reach
an answer in the affirmative, we must
first introduce entropy of coherence.\\

{\it Entropy of coherence or of
incompatibility}. To begin with, we
should notice that the difference
between (14) and (18) lies in {\it
coherence}, which is present in the
former and absent in the latter. One
may wonder if coherence can be given a
precise and general definition.

I suggest to consider the following
quantity as {\it the amount of
coherence or of incompatibility}
between a given observable
$A=\sum_ia_iP^i$ (in the unique
spectral form) and a given quantum
state $\rho$, and call it {\it the
entropy of coherence or of
incompatibility}: $$E_C(A,\rho )\equiv
S(T_A\rho )-S(\rho )\eqno{(19)}$$ (cf
(5)), i. e., the increase of entropy in
ideal nonselective measurement of $A$
in $\rho$.

That the RHS of (19) is always
nonnegative and zero if and only if $A$
and $\rho$ commute (compatibility) was
proved in \cite{Neu} (pp. 380-387) for
complete $A$. That for any state $\rho$
and for any incomplete observable $A$
there always exists a complete one $B$
such that the former is a function of
the latter and such that $T_A\rho
=T_B\rho$ was proved in \cite{FHSpec}
(Theorem 2. there). Hence, the RHS of
(19) is always nonnegative also for
incomplete observables, and it is zero
if and only if $[A,\rho ]=0$. (Namely,
the commutation is sufficient for
$T_A\rho =\rho$, and hence for zero LHS
of (19). On the other hand, the
mentioned zero implies, as stated,
commutation with $B$, and hence also
with $A$.)

Utilizing the mixing property of
entropy, we can rewrite (19) as
$$E_C(A,\rho )=H(A)-\Big(S(\rho )-
\sum_iw_iS(\rho_i)\Big),\eqno{(20)}$$
where $\forall i:\enskip w_i\equiv \Tr
P^i\rho$, $\rho_i\equiv P_i\rho
P_i/w_i$ (for $w_i>0$) and $H(A)\equiv
H(w_i)$ is the mixing entropy, which
is, simultaneously, also the {\it
entropy of the observable} $A$ in
$\rho$.

It was proved in \cite{Lin72} (Theorem
2. there) that, whenever
$S(\rho)<\infty$, the second term on
the RHS of (20) is, in its turn, always
nonnegative, and zero if and only if
$\forall i:\enskip S(\rho_i)=S(\rho)$.
(This condition is satisfied, e. g.,
when $\rho$ and all $\rho_i$ are pure
states, like in the case of measurement
in a pure state.) On the other hand,
the above discussion shows that the
mentioned second term never exceeds the
first; and they are equal if and only
if $[A,\rho ]=0$.

If $A$ is {\it complete} and $\rho$
mixed or pure, then the states $\rho_i$
are pure and $$E_C(A,\rho)= H(A)-S(\rho
).\eqno{(21)}$$ If $\rho$ is {\it pure}
and $A$ is incomplete or complete, the
states $\rho_i$ are again pure, and
$$E_C(A,\rho )=H(A).\eqno{(22)}$$ If
{\it both $A$ is complete}, i. e.,
$\forall i: \enskip
P^i=\ket{i}\bra{i}$, {\it and $\rho$ is
pure}, i. e., $\rho =\ket{\phi
}\bra{\phi }$, then $$E_C(A,\rho
)=H(|f_i|^2),\eqno{(23a)}$$ where
$$\ket{\phi }=\sum_if_i\ket{i}
\eqno{(23b)}$$ is the relevant
expansion.

Now we may face the question if the
twin observables given by (15a) and
(15b) have anything to do with quantum
discord in $\ket{\Phi}_{12}$.\\

{\it Purely quantum information and
coherence in bipartite pure states.-}
The entropy of  coherence of $(A_1
\otimes 1)$ given by (15a) or of its
twin observable $(1\otimes B_2)$ (cf
(15b)) in $\ket{\Phi}_{12}$ (cf (14))
is $H(A)=H(B)=S(1)=S(2)$, which equals
the relative entropy of entanglement
$E_{RE}(\ket{\Phi }_{12})$ or the
quantum discord $\delta(m2\rightarrow
1)$ in this state. In $\rho'_{12}$
given by (18) the analogous coherence
entropies are zero (because
$[(A_1\otimes 1),\rho_{12}']=[(1\otimes
B_2),\rho_{12}']=0$).

Thus, in {\it every pure bipartite
state} $\ket{\Phi }_{12}$ it is not
only true that a pair of twin
observables $A_1$ and $B_2$ "carries"
the quasi-classical part of
correlations, i. e., the one accessible
by subsystem measurement, but it is
also true that {\it the same twin
observables "carry" also the
subsystem-measurement-inaccessible part
of correlations}, i. e., the quantum
entanglement, via the amount of
coherence of any of the twin
observables in the bipartite state.\\

\appendix*\section{1} {\it Proof of
relations (9)} is based on
$\sum_j(Q_2^j)^2=1$, and on $\Tr_2[(
\rho_{12}Q_2^j)Q_2^j]=\Tr_2[Q_2^j
(\rho_{12}Q_2^j]$: $$ T_A\rho_1\equiv
\sum_iP_1^i(\Tr_2\rho_{12})P_1^i
=\sum_iP_1^i[\Tr_2(\sum_jQ_2^j\rho_{12}
Q_2^j)]P_1^i=$$
$$\Tr_2\sum_iP_1^i(\sum_j
Q_2^j\rho_{12}Q_2^j)P_1^i=\Tr_2T_AT_B
\rho_{12}.$$ The second relation in (9)
is proved symmetrically.\\

\appendix*\section{2} {\it Proofs for
the initial claims} in the definition
of twin observables.

As well known, statistical operators,
in particular, the reduced ones, have
purely discrete spectra and their
spectral forms (with distinct
characteristic values) read: $\rho_s=
\sum_kr^s_kQ^k_s$, $s=1,2$. As a
consequence of the commutations in
property (i), one has $\forall
k:\enskip [A_1,Q^k_1]=0,\enskip [B_2,
Q^k_2]=0$. Since the range projectors
$Q_s$ of $\rho_s$ are $Q_s=\sum_kQ^k_s,
\enskip s=1,2$ (all $r^s_k$ are
positive), one has also $[A_1,Q_1]=0,
\enskip [B_2,Q_2]=0$. Hence, the
(topological closures of the) ranges
$\cR(\rho_s)$ \Big($\bar
\cR(\rho_s)=\cR(Q_s)$\Big), $s=1,2$ are
invariant subspaces for $A_1$ and $B_2$
respectively. Further, since also the
characteristic subspaces $\cR(Q^k_1)$
of $\rho_1$ are invariant for $A_1$,
and they are necessarily finite
dimensional (because $\sum_kd^1_kr^1_k=
\Tr \rho_1=1$, where $d^1_k$ is the
multiplicity of $r^1_k$), only discrete
characteristic values of $A_1$ appear
in $\cR(\rho_1)$, and symmetrically for
$B_2$.

Let $\sum_la_lP^l_1$ be the discrete
part of the spectral form (with
distinct characteristic values) of
$A_1$. This operator and $A_1Q_1$ act
equally in $\bar \cR(\rho_1)$. Further,
as already proved, all spectral
projectors of $A_1$ belonging to its
(possible) continuous spectrum are
subprojectors of the null-space
projector $Q_1^{\perp}$. Hence,
$A_1Q_1=\sum_la_l(P^l_1Q_1)$. Omitting
all terms in which $P^l_1Q_1=0$, and
changing the index from $l$ to $i$ in
the remaining sum, one obtains the
spectral form $A_1Q_1=\sum_ia_i
(P_1^iQ_1)$. Obviously, $A_1$ has those
and only those characteristic values
$a_i$ in $\bar \cR(\rho_1)$ for which
$P^i_1Q_1\not= 0$.

On the other hand, the detectable
discrete characteristic values $a_n$ of
$A_1$ in $\rho_{12}$ are those for
which $0<p_n\equiv \Tr (\rho_1P_1^n)$.
One can always write
$\rho_1=\rho_1Q_1$. Therefore, $p_n=\Tr
[\rho_1(P^n_1Q_1)]$. If $P^n_1Q_1=0$,
then $p_n=0$. If $P^n_1Q_1\not= 0$, and
we substitute the spectral form $\rho_1
=\sum_kr^1_kQ^k_1$, then $p_n=\sum_k
r^1_k\Tr (P_1^nQ_1^k)$. (We omit $Q_1$
because $Q_1Q_1^k=Q_1^k$.) Since
$\sum_kP_1^nQ_1^k=P_1^nQ_1$, which is
nonzero by assumption, not all $P_1^n
Q_1^k$ can be zero. The nonzero terms
$r^1_k\Tr (P_1^nQ_1^kP_1^n)$ are
obviously positive. Thus, $p_n>0$, and
$a_n$ is detectable. This bears out the
claim that precisely the detectable
values of $A_1$ in $\rho_{12}$ appear
as its characteristic values in $\bar
\cR(\rho_1)$. (Thus, we can write $i$
instead of $n$ like in the preceding
passage.)

{\it Proof of equivalence of the four
conditions} will be given via the
following closed chain of implications:
(a) $\Rightarrow$ (d) $\Rightarrow$ (b)
$\Rightarrow$ (c) $\Rightarrow$ (a).

{\it LINK} (a) $\Rightarrow$ (d).

Let $$\rho_{12}=\sum_kw_k
\ket{\Phi}^k_{12}\bra{\Phi}^k_{12}
\eqno{(A.1)}$$ be a (convex linear)
decomposition of $\rho_{12}$ into ray
projectors. (For instance, the
$\ket{\Phi}^k_{12}$ can be the
characteristic state vectors of
$\rho_{12}$.) If a projector $E$ is
probability-one in $\rho_{12}$, then so
is it in each $\ket{\Phi}^k_{12}$ (as
seen from $1=\Tr (\rho_{12}E)=\sum_kw_k
\Tr
(\ket{\Phi}^k_{12}\bra{\Phi}^k_{12}E)$
and $\sum_kw_k=1$). Further, $$1=
\bra{\Phi}^k_{12}E\ket{\Phi}^k_{12}\enskip
\Rightarrow \enskip 0=\bra{\Phi}^k_{12}
E^{\perp}\ket{\Phi}^k_{12}\enskip
\Rightarrow $$
$$||E^{\perp}\ket{\Phi}^k_{12}||^2=0
\enskip \Rightarrow \enskip E^{\perp}
\ket{\Phi}^k_{12}=0\enskip \Rightarrow
\enskip
E\ket{\Phi}^k_{12}=\ket{\Phi}^k_{12}.$$
The sum $\sum_iP_1^i$ ($\sum_iP_2^i$)
of all detectable values of $A_1$
($B_2$) is a probability-one projector
in $\rho_{12}$. Therefore, $$\forall
k:\quad \ket{\Phi}^k_{12}=(\sum_iP_1^i)
\ket{\Phi}^k_{12}
=(\sum_iP_2^i)\ket{\Phi}^k_{12},$$ and
$$\ket{\Phi}^k_{12}=(\sum_iP_1^i)
(\sum_iP_2^i)\ket{\Phi}^k_{12}=
\sum_{ii'}P_1^iP_2^{i'}\ket{\Phi}^k_{12}
.\eqno{(A.2)}$$

Assuming the validity of condition (a),
and utilizing (A.1), we have $$i\not=
i'\quad \Rightarrow \quad 0=p_{ii'}
\equiv \Tr \rho_{12}P_1^iP_2^{i'}$$ $$=
\sum_kw_k\bra{\Phi}^k_{12}P_1^iP_2^{i'}
\ket{\Phi}^k_{12}.$$

Since $\forall k:\enskip w_k>0$, the
second factor in each term in this sum,
generally nonnegative, must be zero.
This implies, by making use of the
definiteness of the norm as above, that
for distinct $i$ and $i'$ $$\forall
k:\quad
P_1^iP_2^{i'}\ket{\Phi}^k_{12}=0.
\eqno{(A.3)}$$ Relations (A.2) and
(A.3) imply $$\forall k,i: \quad P_1^i
\ket{\Phi}^k_{12}=P_1^iP_2^i
\ket{\Phi}^k_{12}=P_2^i
\ket{\Phi}^k_{12}.\eqno{(A.4)}$$
Relation (A.4) in conjunction with
(A.1) finally gives condition (d).

{\it LINK} (d) $\Rightarrow$ (b)

Making use of condition (d) and its
adjoint in the LHS of condition (b),
this condition is immediately derived.

{\it LINK} (b) $\Rightarrow$ (c)

The LHS of condition (c) can be
rewritten as $$\forall i:\quad p_i^{-1}
\Tr P_1^i(P_2^i\rho_{12}P_2^i)P_1^i.$$
If one utilizes condition (b), this
expression becomes $p_i^{-1}p_i$, i.
e., condition (c) follows.

{\it LINK} (c) $\Rightarrow$ (a)

Let us return to the argument given in
the proof of the link \Big((a)
$\Rightarrow$ (d)\Big), and to (A.1).
It was shown that a probability-one
projector $E$ in $\rho_{12}$ is such an
event also in each $\ket{\Phi}^k_{12}$,
and $\forall k:\enskip
E\ket{\Phi}^k_{12}=\ket{\Phi}^k_{12}$.
Then, (A.1) implies $$E\rho_{12}=
\rho_{12}.\eqno(A.5)$$

Assuming the validity of (c), $P_2^i$
is a probability-one projector in
$\rho_2(P_1^i)$, hence, on account of
the adjoint of (A.5), one has $$\rho_2
(P_1^i)=\rho_2
(P_1^i)P_2^i.\eqno{(A.6)}$$ The LHS of
condition (a), due to (A.6),  implies
$$p_{ii'}\equiv \Tr (\rho_{12}P_1^iP_2
^{i'})=p_i\Tr [\rho_2(P_1^i)P_2^{i'}]$$
$$= p_i\Tr \Big[\Big(\rho_2(P_1^i)P_2^i
\Big)P_2^{i'}\Big]=\delta_{i,i'}p_i.$$
Thus, (a) is derived.

{\it Proof of the stronger algebraic
relation}.

Since
$\rho_{12}=(\sum_iP_1^i)\rho_{12}$, one
has
$A_1\rho_{12}=(\sum_ia_iP_1^i)\rho_{12}$.
Assuming then property (iv), i. e.,
$\forall i:\enskip a_i=b_i$, and
utilizing condition (d), one further
obtains $$A_1\rho_{12}=
(\sum_ib_iP_2^i)\rho_{12}=B_2\rho_{12}.$$
The last equality is due to the fact
that for the second subsystem one has
the symmetric argument. Thus, the
stronger algebraic relation is
derived.\hfill $\Box$\\

\end{document}